# Transistor behavior via Au clusters etched from electrodes in an acidic gating solution: metal nanoparticles mimicking conducting polymers

Jacob E. Grose*, Abhay N. Pasupathy, and D. C. Ralph

Laboratory of Atomic and Solid State Physics, Cornell University, Ithaca, NY USA

14853

Burak Ulgut* and Héctor D. Abruña

Department of Chemistry and Chemical Biology, Cornell University, Ithaca, NY USA

14853

* These authors contributed equally to this work.

We report that the electrical conductance between closely-spaced gold electrodes in acid solution can be turned from off [insulating; I] to on [conducting; C] to off again by monotonically sweeping a gate voltage applied to the solution. We propose that this ICI transistor action is due to an electrochemical process dependent on nanoparticles etched from the surface of the gold electrodes. These measurements mimic closely the characteristics of nanoscale acid-gated polyaniline transistors, so that researchers should guard against misinterpreting this effect in future molecular-electronics experiments.

85.35.-p

In recent years, improved nanofabrication techniques and chemical-assembly procedures have led to promising initial results in the field of molecular electronics.[1,2] For future research in this field, conjugated conducting polymers such as polyaniline (PANI) represent a promising class of materials because they have peak conductivities near those of conventional metals, and their electrical properties can be controlled by gating.[3] Pioneering work on PANI thin-film transistors with lateral dimensions on the micron scale was performed by Paul, Ricco and Wrighton (PRW).[4] More recently, a variety of devices using PANI have been made on smaller scales, employing nanometer-scale electropolymerization of aniline,[5] the formation of electrospun nanofibers,[3] and measurements of polyaniline nanowires using STM[6] and nanoelectrodes.[7] Our group has been involved in making PANI based transistors between metal (gold) electrodes spaced by just 1-2 nm in an effort to access the regime of single polymer chains. In the course of this work, we have observed a behavior that closely mimics the transistor properties to be expected from small numbers of PANI strands, even when our devices contain no molecular species. While we do not challenge the conclusions of any previous studies, given that many artifacts which can occur in the field of molecular electronics are not well characterized or understood, we report here an investigation of this effect. We propose that the observed transistor behavior arises as a result of an electrochemical process in which gold nanoparticles etched from closely-spaced source and drain gold electrodes shuttle electrons between the electrodes to give a measurable conductance.

The electrical properties of transistors based on electropolymerized thin films of PANI are nicely illustrated by the results of PRW.[4] When gold electrodes are coated with a PANI film in acid solution and one sweeps a gate voltage applied to the solution, the conductance of the film initially increases by a factor of $10^6$ and then decreases back to a negligible value as the voltage is further increased. The mechanism behind this modulation in conductance relies on the fact that PANI has three distinct structural forms, only one of which is conducting in an acid solution. As the gate potential is swept, the potential between the solution and the drain electrode forces the PANI film to change from the insulating leucoemeraldine structure to the conducting emeraldine structure and then subsequently to the insulating pernigraniline structure.[8]

Our devices are composed of three gold electrodes as shown in Fig. 1: a macroscopic wire that serves as the gate electrode and two nanoscale electrodes, supported on a silicon substrate with a thick oxide layer, that serve as the source and drain. The source and drain electrodes were initially fabricated as a continuous wire using a combination of electron-beam lithography and photolithography, and then this wire was broken into separate contacts using electromigration.[9] The resulting gap is typically a few nanometers wide at room temperature. Although this gap can be bridged by polyaniline via electropolymerization, in this paper we focus on the properties of devices with bare gold electrodes. The macroscopic gold wire that acts as the gate electrode was placed in a glass micropipette with a tip diameter of around 10 microns filled with an aqueous solution of 0.5 M $HClO_4$. The pipette was positioned over a silicon chip containing the source and drain electrodes, after the chip was pre-cleaned with an oxygen plasma. As the chip was brought up to the pipette tip, the hydrophilic surface drew the solution from the micropipette over the chip's surface, forming a small drop of solution over the electrodes. The drop remained connected at all times to the reservoir of solution in the pipette and thus was in contact with the macroscopic gate electrode which controlled its potential.

The drop of acid solution covered not only the thinnest regions of the source and drain electrodes, but also part of the 15-$\mu$m-wide contacts (see Fig. 1).

Note that our setup does not include a reference electrode, and the convention of our gate voltage is the opposite of that usually adopted in electrochemical measurements. The amount of electrolyte on our chip is too small to allow the use of a conventional reference electrode. We have many gold connections on each chip so that we must employ a micropipette to localize the electrolyte about an individual device. Otherwise the electrolyte potential could not be controlled by the gate electrode and the gate electrode alone. We have considered the possibility of using lithography to define a silver reference wire on the chip, but this would introduce the risk electroplating silver onto the gold break junctions. However, as explained below the absence of a reference electrode does not compromise the interpretation of our data.

Experimental current-potential curves for bare gold source and drain electrodes in contact with aqueous 0.5 M $HClO_4$ are shown in Fig. 2. The data were taken during source voltage sweeps from -600 mV to 600 mV at 50 mV/s at a constant $V_g$ value. As $V_g$ is varied with $V_s$ near 0, we find a peak in the electrical conductance near $V_g$ = -0.8 V (although this value varies somewhat from device to device), with negligible conductances for lower and higher $V_g$. We observe positive drain current for positive $V_s$ over the region -0.8 V < $V_g$ < -0.4 V and negative drain current for negative $V_s$ over the region -1.4 V < $V_g$ < -0.8 V. Drain current levels are typically 1 nA at $V_{sd}$ = 600 mV, with transconductances of 3 nS. We have observed similar characteristics for over 30 different devices. The current levels are not particularly sensitive to the spacing of the source and drain electrodes. We measure currents of hundreds of pA for $V_{sd}$ = 600 mV with electrode spacings as wide as 0.5 $\mu$m. Since the drop of acid solution bridges the full 10-$\mu$m gap between our large photolithographically-defined contacts, it is likely that a significant amount of current flows through the solution over this long distance and not just through the nm-scale gap between the source and drain electrodes.

For purposes of comparison, in Fig. 3 we plot the current-voltage characteristics for a 100-nm-thick PANI film grown over a nanoscale gap in our electrodes via electropolymerization of the monomer; the same method employed by PRW.[4] Since the drain current levels are 4 orders of magnitude greater in this PANI device than in devices without the polymer (i.e. Fig 2), we can conclude that in Fig. 3 the current flows through the PANI film. However, the dependence of the current on $V_s$ and $V_g$ is qualitatively similar in both cases. In particular, the current in the devices without any PANI film present turns from off to on and then off again as a function of gate voltage just like the current in devices with a PANI film. While we do not doubt the interpretation of any PANI studies on the micron scale, the current levels in future experiments that reach down to the regime of single PANI chains should be comparable to the currents we observe in devices without any molecules. Therefore, there is a distinct danger of misinterpretation.

As a first step toward understanding the mechanism of our transistor action, consider the gate current shown in Fig. 2b for the device without a PANI film. This plot has three distinct regions of non-zero current, regions which are common to many metal electrodes in aqueous acid electrolytes.[10] The steady-state currents at the corners of the graph are due to $H_2$ and $O_2$ evolution at the source electrode. The diagonal ridge in the center of the graph is a transient current due primarily to oxidation and reduction of the source electrode surface. We can be confident of this interpretation, despite the absence of a reference electrode, because these are the only reactions known to take place on gold surfaces in aqueous acid. The oxidation and reduction of the drain electrode will not induce any measurable current at the gate electrode because the gate current should be proportional[11] to the sweep rate of $V_g$ with respect to $V_d$, and in our experiment neither the drain nor the gate electrodes are swept, but instead held at constant potentials relative to one another. Since the diagonal ridge in the gate current occurs at the same potentials as the diagonal threshold in the drain current separating the regions of high and low current (Fig. 2a), it is clear that whatever process is causing the drain current is, in some way, related to the oxidation and reduction of the gold source electrode.

In Fig. 4 we have replaced the perchloric acid (0.5 M $HClO_4$) with an aqueous solution of sodium perchlorate ($NaClO_4$) at the same ionic strength (0.5 M concentration) but of essentially neutral pH, and we performed the same measurements as were shown for bare gold electrodes in Fig. 2. Note that the characteristic transistor behavior of the drain current in acid solution (Fig. 2a) is completely absent in Fig. 4, despite the much wider window of gate potential (electrochemical window) displayed, extending from $O_2$ evolution on the left to $H_2$ evolution on the right. This demonstrates that neither the perchlorate ion nor the deionized water itself is responsible for the transistor characteristics observed, since these species are present in both the acid and salt solutions.

We can also rule out the possibility that the current-carrying species in our devices might be individual gold ions because the reduction potentials for the most stable gold ions Au(I) and Au(III) are higher than the potential needed to oxidize $H_2O$ in a pH = 0 aqueous solution, placing them well outside of our measurement window.[11] At the highly negative $V_g$ potentials at which gold ions would be oxidized on the drain electrode, any drain currents due to these ions would be masked by much stronger currents due to the oxidation of $H_2O$.

We propose that the transistor characteristics we observe in devices with bare gold electrodes result from the formation of gold nanoparticles etched from the source and drain electrodes by the acid electrolyte when the source voltage is swept. Without acid to etch the electrodes, no transistor characteristics are observed. Our electrodes are clearly etched during the measurement process, as shown by scanning electron microscope (SEM) images (Fig. 5). In previous STM measurements on Au(111) surfaces undergoing electrochemical cycling in 1 M sulfuric acid, Nieto *et al.* inferred the formation of gold clusters with diameters on the order of a few nanometers.[12] To verify the presence of gold nanoparticles in our samples, we emulated our device setup on a larger scale using macroscopic gold wires, and we applied potentials similar to those used for the nanoscale devices. The resulting acid solution was then placed on a silicon wafer and evaporated. Clusters of gold nanoparticles could be seen using SEM (Fig. 6). We confirmed that the nanoparticles were gold using EDX spectroscopy on the same sample.

We suggest that the underlying mechanism by which the current flows in our devices is the same mechanism that underlies scanning electrochemical microscopy (SECM) in feedback mode.[11] The potential of the solution can be tuned using the gate voltage to a value such that a small bias between the source and drain can allow the surfaces of gold nanoparticles to be oxidized at one electrode and reduced at the other, and as a result the nanoparticles will shuttle charge between the electrodes to generate a steady-state current. Current will flow whenever the applied gate voltage causes the redox potential to lie between the source and drain potentials. This provides an explanation for the slopes of the lines separating the regions of high drain current from the regions of low drain current in Fig. 2a. If we assume that the capacitance between the gate electrode and the solution is much greater than the capacitance between the other electrodes and the solution, the slope of the diagonal line should be $V_s/V_g = C_g/(C_g + C_d) \approx 1$, and the slope of the vertical line should be $V_s/V_g = -C_g/C_s \approx -\infty$, in agreement with the data. However, in standard feedback-mode SECM one expects to be able to turn a current from off to on by sweeping $V_g$, but not to turn the current from off to on to off again as in our devices. This is because in most SECM applications there is a uniform solution concentration of electroactive species far away from the source and the drain. This concentration is generally high enough that, reactions at the electrodes do not significantly perturb the bulk concentration.

The unique behavior of our device can stem from the fact that since the gold nanoparticles are etched from the electrodes themselves, the local concentration of particles between the two leads can be much greater than the concentration in the rest of the solution. This means that if all of the nanoparticles begin in the oxidized state, it is plausible that most could be reduced rather quickly if the potential of the solution were shifted so that reduction occurred on both the source and the drain. This permits the conductance to start near zero when oxidation occurs on both electrodes, rise to some finite value when oxidation occurs on one electrode while reduction occurs on the other, and then drop back to zero when reduction occurs on both electrodes.

To check whether this model is reasonable, we have performed an order-of-magnitude estimate for the current. The maximum steady-state current density one can expect from electron transfer through a diffusing redox species is given[13] by $J = neDC/d$, where $n$ is the number of electrons transferred per nanoparticle, $D$ is the diffusion coefficient of the redox species which is assumed to be the same for both the reduced and oxidized species, $C$ is the equilibrium concentration of the majority species assumed to be constant between the electrodes, and $d$ is the distance between the electrodes. In our experiment we expect that we are forming gold hydroxide (as opposed to gold oxide)[12] on the nanoparticle surface at one electrode and removing it at the other. Particles carrying a surface hydroxide are still electrically neutral, despite the electron-transfer reaction. Therefore, Coulomb charging effects need not limit the oxidation process, and it is likely that the majority of the nanoparticle surface can be oxidized simultaneously. For particles 10 nm in diameter, this suggests that the charge transferred per particle is likely to be on the order of $n = 1000$.

If we take the maximum current density in our devices to be 1 nA spread over the maximum cross-sectional area covered by the drop, 0.24 $mm^2$, and we take d to be 10 μm (see Fig. 1) and D to be $10^{-6}$ cm$^2$/s,[14] we find C to be on the order of 1 mM. Although this concentration seems plausible, further work, such as an experimental measurement or a Monte Carlo simulation, would be necessary in order to confirm this value.

In summary, we find that bare gold electrodes gated by an acid electrolyte can exhibit transistor-like electrical characteristics that closely mimic the properties of polyaniline transistors. We have proposed a mechanism in which gold nanoparticles etched from the source and drain electrodes shuttle charge between these electrodes through cycles of redox reactions. This process is analogous to the principle of an SECM operating in feedback mode, after taking into account a large local concentration of nanoparticles. We hope that an understanding of this mechanism will prevent potential future misinterpretations of nanoscale experiments on acid-gated molecular transistors.

We acknowledge support from the NSF/MRSEC program through the Cornell Center for Materials Research, ONR (N00173-03-1-G011), and ARO (DAAD19-01-1-0541). We also acknowledge use of the NSF-supported Cornell Nanofabrication Facility/NNIN. We thank Dr. Jiwoong Park and Prof. Paul McEuen for providing initial directions for the project, Prof. Alejandro Arvía for helpful discussions and Mr. Çaglar Çitir for his help in preparation of the artwork.

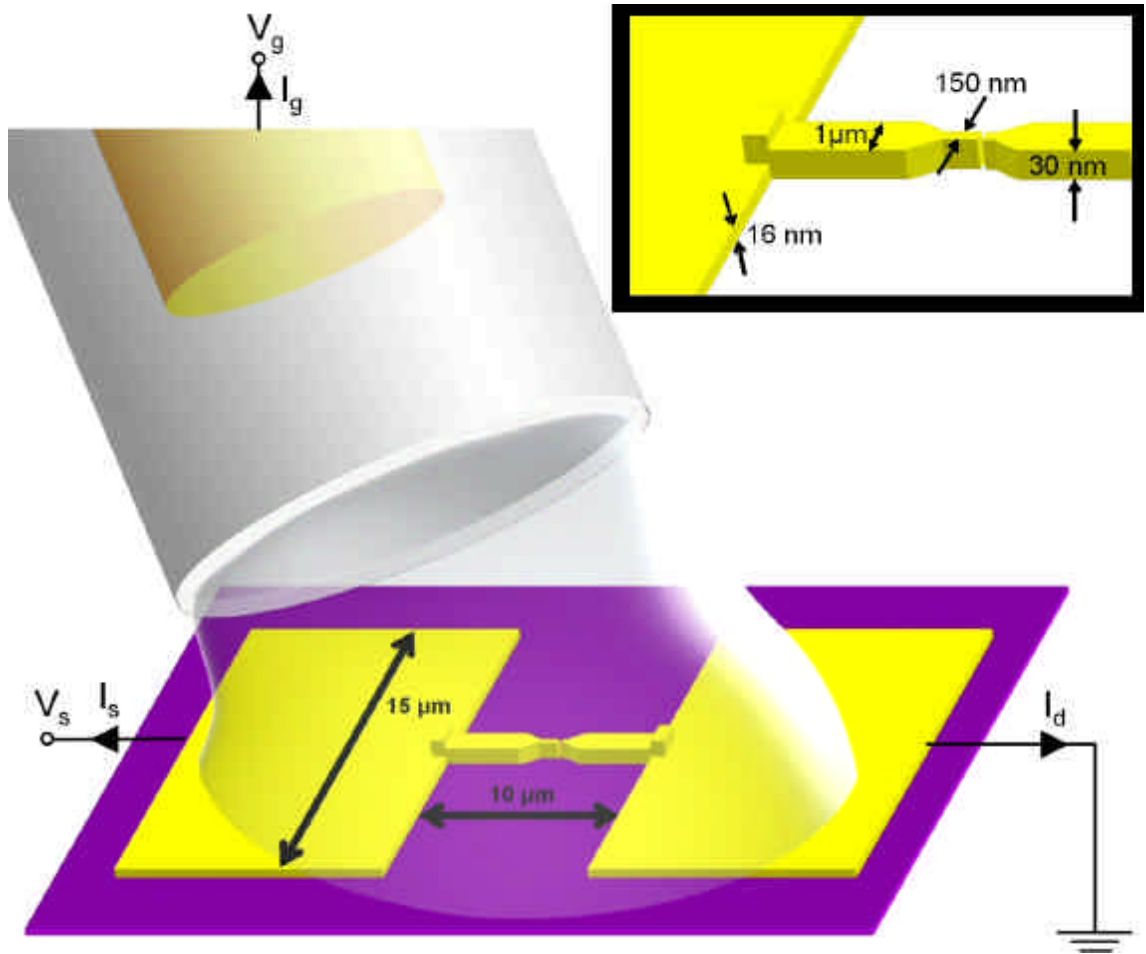

**Figure 1. Schematic of the device geometry. Dimensions of the smaller features of the lithography are given in the inset. The electrolyte is brought into contact with the junction via a glass pipette tip a few microns in diameter. Voltages are applied to the source and gate electrodes while the drain is attached to ground. Current directions are defined as indicated.**

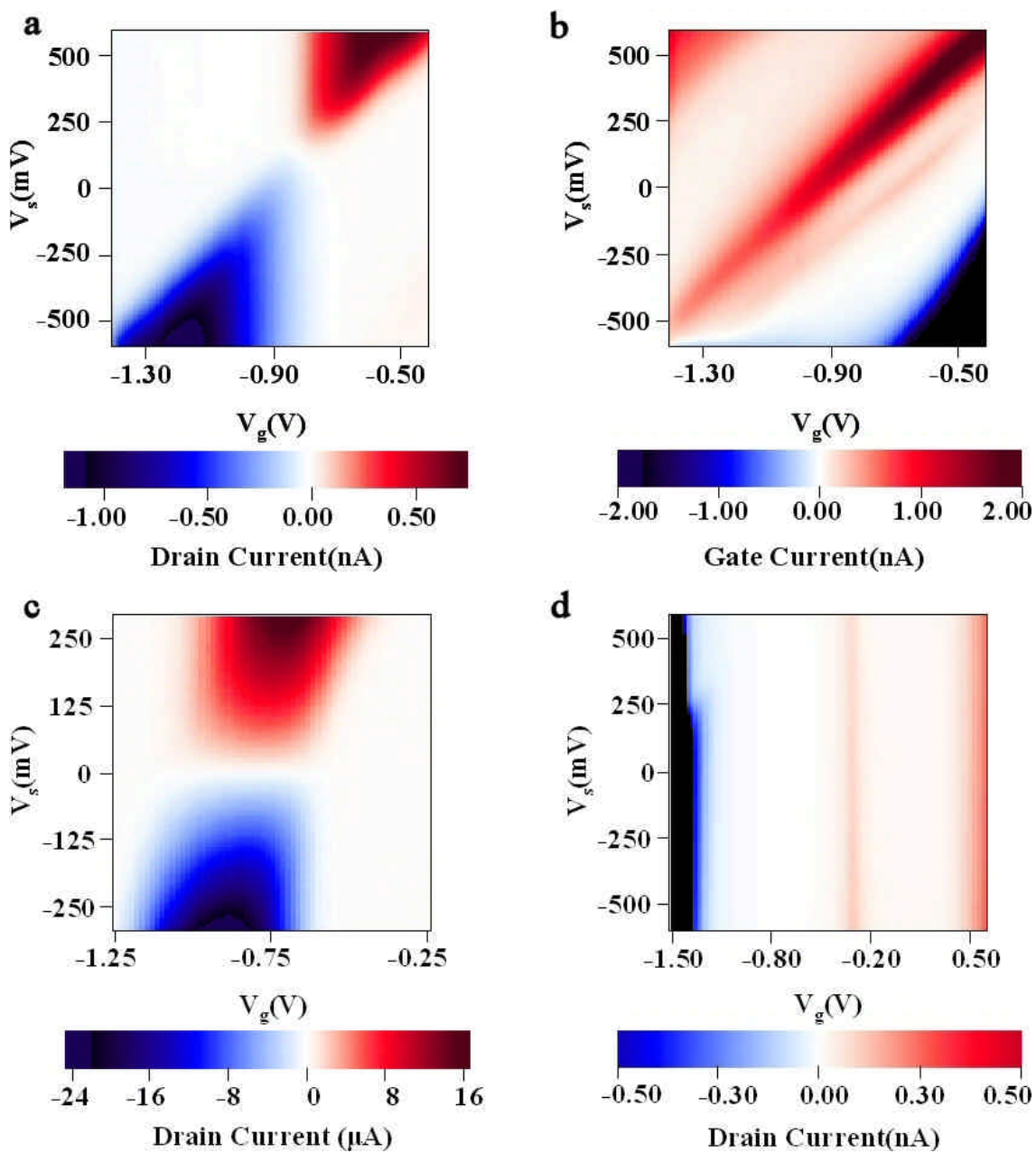

**Figure 2.** (a) Drain current and (b) gate current measured in a device with bare gold electrodes under aqueous 0.5 M HClO4. The diagonal ridge of positive gate current in (b) corresponds to the diagonal threshold for the drain current to turn on in (a). In this region, the Au source electrode undergoes oxidation. (c) Drain current for a device with a 100-nm-thick PANI film grown between the source and drain

electrodes, measured under pH 1 sulfuric acid solution. The pattern in which the drain current as a function of gate voltage goes from off to on to off is similar to the device with bare electrodes in (a), although the transition region is wider and the current scale is four orders of magnitude larger. (d) Drain current for a device with bare gold electrodes under a neutral-pH solution of aqueous 0.5 M NaClO4, shown for a larger range of gate voltage than in (a). Note that the transistor characteristics seen in (a) are absent in (d) throughout the entire region of bias extending from oxygen evolution (blue region on the left) to hydrogen evolution (red region on the right).

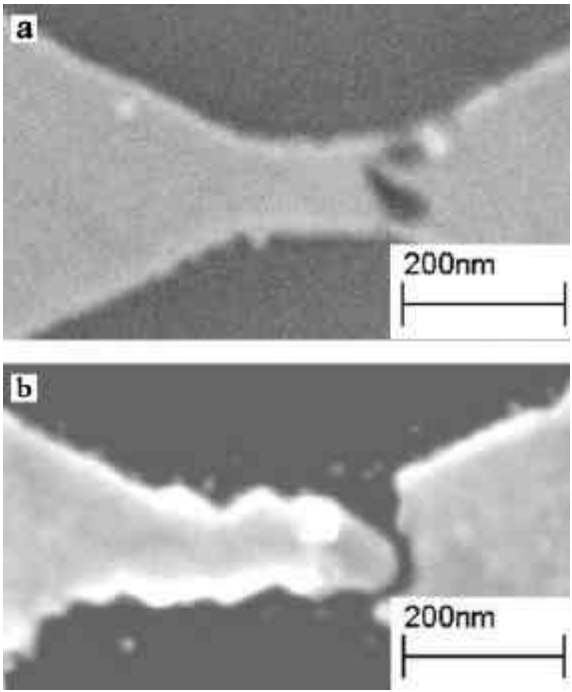

**Figure 3. SEM images of two Au junctions (a) after being broken in air by electromigration and (b) after being broken in air by electromigration, then placed under aqueous 0.5 M HClO$_4$ at constant V$_g$ in the region of positive current with V$_s$ (left electrode) cycled between –600 mV and 600 mV at 50 mV/s for 10 minutes, and then dried. Note that the gold wire in the second junction has been etched by this process.**

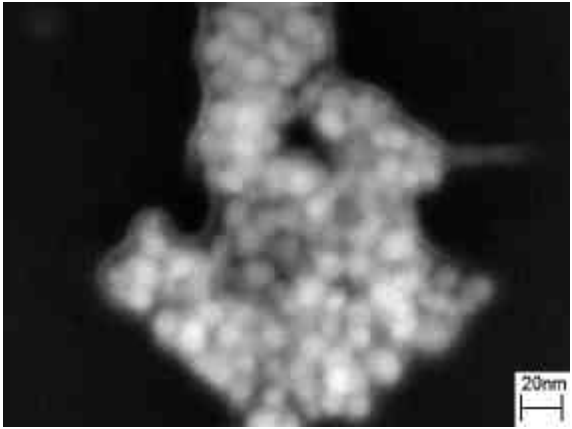

**Figure 4. SEM image of a cluster of Au nanoparticles collected from aqueous 0.5 M HClO$_4$ solution after this solution was used to etch macroscopic gold wires via potential sweeps similar to those used to measure our devices. The solution was dried on a Si substrate before imaging.**